\documentclass[11pt,a4paper,DIV=11,numbers=noenddot,parskip=half]{scrartcl}
\pdfoutput=1

\usepackage[T1]{fontenc}
\usepackage{lmodern}
\usepackage[utf8]{inputenc}
\usepackage{amsmath}
\usepackage{amssymb}
\usepackage{graphicx}
\usepackage{enumitem}
\usepackage[absolute]{textpos}
\usepackage[noadjust]{cite}
\usepackage[affil-it]{authblk}
\usepackage{xcolor}
\usepackage{tabularx}
\usepackage{booktabs}
\usepackage{multirow}
\usepackage{listings}
\usepackage{bold-extra}
\usepackage{float}
\usepackage[font=small,labelfont=bf,format=plain,margin=0.05\textwidth]{caption}
\usepackage[pdftitle={Kira 3: integral reduction with efficient seeding and optimized equation selection},
  pdfauthor={Fabian Lange, Johann Usovitsch, Zihao Wu},
  pdfkeywords={Feynman diagrams, multi-loop Feynman integrals, integral reduction, computer algebra},
  bookmarks=true, linktocpage,
  colorlinks=true, allbordercolors=white, allcolors=blue]{hyperref}
\usepackage[capitalize]{cleveref}
\usepackage{orcidlink}
\usepackage{acronym}
\usepackage[labelformat=simple]{subcaption}

 \newcounter{notecount}

\newcommand{\citere}[1]{Ref.~\cite{#1}}
\newcommand{\citeres}[1]{Refs.~\cite{#1}}

\newcommand{\code}[1]{\texttt{#1}}
\newcommand*{\kira}{\code{Kira}}
\newcommand*{\neatibp}{\code{NeatIBP}}
\newcommand*{\ratracer}{\code{Ratracer}}

\newcommand*{\kirathr}{\code{Kira\;3}}
\newcommand*{\kiratwothr}{\code{Kira\;2.3}}
\newcommand*{\pyred}{\code{pyRed}}
\newcommand*{\firefly}{\code{FireFly}}

\newcommand*{\fermat}{\code{Fermat}}

\newcommand{\pheaderline}{{\footnotesize ZU-TH 39/25\\HU-EP-25/17-RTG}}

\acrodef{ibp}[IBP]{integration-by-parts}
\acrodef{Ibp}[IBP]{Integration-by-parts}

\title{Kira 3: integral reduction with efficient seeding and optimized equation selection}

\author[,a,b]{Fabian Lange\,\orcidlink{0000-0001-8531-5148}%
  \thanks{E-mail: \href{mailto:fabian.lange@physik.uzh.ch}{fabian.lange@physik.uzh.ch}}}

\author[,c]{Johann Usovitsch\,%
\orcidlink{0000-0002-3542-2786}%
  \thanks{E-mail: \href{mailto:johann.usovitsch@hu-berlin.de}{johann.usovitsch@hu-berlin.de}}}

\author[,d]{Zihao Wu\,\orcidlink{0000-0003-3561-5403}%
  \thanks{E-mail: \href{mailto:wuzihao@mail.ustc.edu.cn}{wuzihao@mail.ustc.edu.cn}}}

\affil[a]{Physik-Institut, Universität Zürich, Winterthurerstrasse 190, 8057 Zürich, Switzerland}
\affil[b]{PSI Center for Neutron and Muon Sciences, Forschungsstrasse 111, 5232 Villigen PSI, Switzerland}
\affil[c]{Institut f\"ur Physik und IRIS Adlershof, Humboldt–Universität zu Berlin, 10099 Berlin, Germany}
\affil[d]{School of Fundamental Physics and Mathematical Sciences, Hangzhou Institute for Advanced Study, UCAS, 310024 Hangzhou, China}

\date{}

\begin{document}

\maketitle
\thispagestyle{empty}

\begin{abstract}
  We present version \code{3} of \kira{}, a Feynman integral reduction program for high-precision calculations in quantum field theory and gravitational-wave physics.
  Building on previous versions, \kirathr{} introduces optimized seeding and equation selection algorithms, significantly improving performance for multi-loop and multi-scale problems.
  New features include convenient numerical sampling, symbolic integration-by-parts reductions, and support for user-defined additional relations.
  We demonstrate its capabilities through benchmarks on two- and three-loop topologies, showcasing up to two orders of magnitude improvement over \kiratwothr{}. \kirathr{} is publicly available and poised to support ambitious projects in particle physics and beyond.
\end{abstract}


\begin{textblock*}{11em}(\textwidth,17mm)
\raggedright\noindent
\pheaderline
\end{textblock*}

\newpage

\section*{NEW VERSION PROGRAM SUMMARY}

\textit{Program title:}
\kira{}

\textit{Developer's repository link:}
\url{https://gitlab.com/kira-pyred/kira}

\textit{Licensing provisions:}
GNU General Public License 3 (GPL)

\textit{Programming language:}
\texttt{C++}

\textit{Journal Reference of previous version:}\\
P.~Maierhöfer, J.~Usovitsch and P.~Uwer,
\emph{{Kira—A Feynman integral reduction program}},
\href{https://doi.org/10.1016/j.cpc.2018.04.012}{\emph{Comput.\ Phys.\ Commun.}~{\bfseries 230} (2018) 99}
[\href{https://arxiv.org/abs/1705.05610}{{\ttfamily 1705.05610}}].\\
J.~Klappert, F.~Lange, P.~Maierhöfer and J.~Usovitsch,
\emph{{Integral reduction with Kira 2.0 and finite field methods}},
\href{https://doi.org/10.1016/j.cpc.2021.108024}{\emph{Comput.\ Phys.\ Commun.}~{\bfseries 266} (2021) 108024}
[\href{https://arxiv.org/abs/2008.06494}{{\ttfamily 2008.06494}}].

\textit{Does the new version supersede the previous version?:}
Yes.

\textit{Reasons for the new version:}
Improved algorithms with significant performance gains for all problems and new features.

\textit{Summary of revisions:}
The primary new feature is an improved seeding and selection of equations.
Further improvements include the expanded support for numerical integration-by-parts applications, symbolic integration-by-parts reductions, and support for user-defined additional relations.

\textit{Nature of problem:}
The reduction of Feynman integrals to a smaller set of master integrals is a central building block for high-precision calculations of observables in theoretical particle and gravitational-wave physics.
Furthermore, the reduction is a key ingredient in many methods to calculate the master integrals themselves.

\textit{Solution method:}
\kira{} generates a system of equations using integration-by-parts~[1,2], Lorentz-invariance~[3], and symmetry relations.
It eliminates linearly dependent equations and identifies master integrals by solving the system over a finite field~[4] and then solves the system of equation with Laporta's algorithm~[5].
Two solution methods are available: since version \code{1.0}, \kira{} algebraically solves the system using \fermat{}~[6]; since version \code{2.0}, it numerically solves the system multiple times over finite fields, reconstructing master integral coefficients with \firefly{}~[7,8].
Both approaches extend to any homogeneous linear system.
New in this version, an optimized algorithm for seeding and selecting integration-by-parts identities demonstrates that a small subset of prior equations suffices for a full reduction.

\textit{References:}

[1] F.~V. Tkachov, \emph{{A theorem on analytical calculability of 4-loop
  renormalization group functions}},
  \href{https://doi.org/10.1016/0370-2693(81)90288-4}{\emph{Phys.\ Lett.\ B}
  {\bfseries 100} (1981) 65}.

[2] K.~G. Chetyrkin and F.~V. Tkachov, \emph{{Integration by parts: The algorithm
  to calculate $\beta$-functions in 4 loops}},
  \href{https://doi.org/10.1016/0550-3213(81)90199-1}{\emph{Nucl.\ Phys.}~{\bfseries B192}
  (1981) 159}.

[3] T.~Gehrmann and E.~Remiddi, \emph{{Differential equations for two-loop
  four-point functions}},
  \href{https://doi.org/10.1016/S0550-3213(00)00223-6}{\emph{Nucl.\ Phys.}~{\bfseries B580}
  (2000) 485}
  [\href{https://arxiv.org/abs/hep-ph/9912329}{{\ttfamily hep-ph/9912329}}].

[4] P.~Kant, \emph{{Finding linear dependencies in integration-by-parts equations:
  A Monte Carlo approach}},
  \href{https://doi.org/10.1016/j.cpc.2014.01.017}{\emph{Comput. Phys. Commun.}
  {\bfseries 185} (2014) 1473}
  [\href{https://arxiv.org/abs/1309.7287}{{\ttfamily 1309.7287}}].

[5] S.~Laporta, \emph{{High precision calculation of multiloop Feynman integrals by
  difference equations}},
  \href{https://doi.org/10.1142/S0217751X00002159}{\emph{Int. J. Mod. Phys. A}
  {\bfseries 15} (2000) 5087}
  [\href{https://arxiv.org/abs/hep-ph/0102033}{{\ttfamily hep-ph/0102033}}].

[6] R.~H. Lewis, \emph{{Computer Algebra System Fermat}},
  \href{https://home.bway.net/lewis}{https://home.bway.net/lewis}.

[7] J.~Klappert and F.~Lange, \emph{{Reconstructing rational functions with
  FireFly}}, \href{https://doi.org/10.1016/j.cpc.2019.106951}{\emph{Comput.\
  Phys.\ Commun.}~{\bfseries 247} (2020) 106951}
  [\href{https://arxiv.org/abs/1904.00009}{{\ttfamily 1904.00009}}].

[8] J.~Klappert, S.~Y. Klein and F.~Lange, \emph{{Interpolation of dense and
  sparse rational functions and other improvements in FireFly}},
  \href{https://doi.org/10.1016/j.cpc.2021.107968}{\emph{Comput.\
  Phys.\ Commun.}~{\bfseries 264} (2021) 107968}
  \href{https://arxiv.org/abs/2004.01463}{{\ttfamily 2004.01463}}.

\newpage

\pagenumbering{Alph}
\clearpage

\tableofcontents
\pagenumbering{roman}

\clearpage

\pagenumbering{arabic}

\section{Introduction}

Feynman integrals are central objects for high-precision predictions for both particle physics experiments based on perturbative quantum field theory as well as for scattering events of cosmological objects like black holes which provide input for gravitational-wave studies.
Calculations first generate amplitudes or angles consisting of tensor Feynman integrals which are then reduced to scalar Feynman integrals with the help of tensor reduction techniques.
State-of-the-art calculations typically result in at least tens of thousands if not millions of scalar integrals whose direct calculations is infeasible in most cases.
\ac{Ibp} identities~\cite{Tkachov:1981wb,Chetyrkin:1981qh} and the Laporta algorithm~\cite{Laporta:2000dsw} allow us to reduce this number to a significantly smaller set of master integrals.
This is implemented in tools like \texttt{AIR}~\cite{Anastasiou:2004vj}, \texttt{FIRE}~\cite{Smirnov:2008iw,Smirnov:2013dia,Smirnov:2014hma,Smirnov:2019qkx,Smirnov:2023yhb}, \texttt{Reduze}~\cite{Studerus:2009ye,vonManteuffel:2012np}, \kira{}~\cite{Maierhofer:2017gsa,Klappert:2020nbg}, \textsc{FiniteFlow}~\cite{Peraro:2019svx} with \texttt{LiteRed}~\cite{Lee:2012cn,Lee:2013mka}, and \texttt{Blade}~\cite{Guan:2024byi}.

The ever increasing need of higher precision leads to an increasing complexity both in terms of loop orders and external legs and, thus, requires constant development of new and improved \ac{ibp} reduction strategies.
The field of Feynman integral reductions is constantly evolving and new techniques are explored and applied to integral reduction problems, e.g.\ syzygy equations/algebraic geometry~\cite{Gluza:2010ws,Schabinger:2011dz,Ita:2015tya,Chen:2015lyz,Larsen:2015ped,Bohm:2017qme,Bohm:2018bdy,Bendle:2019csk,vonManteuffel:2020vjv,Wu:2023upw,Wu:2025aeg}, intersection numbers~\cite{Mastrolia:2018uzb,Frellesvig:2019kgj,Frellesvig:2019uqt,Weinzierl:2020xyy,Frellesvig:2020qot,Chestnov:2022alh,Chestnov:2022xsy,Fontana:2023amt,Brunello:2023rpq,Jiang:2023oyq,Crisanti:2024onv,Brunello:2024tqf,Lu:2024dsb}, finite field and interpolation techniques~\cite{Kauers:2008zz,Kant:2013vta,vonManteuffel:2014ixa,Peraro:2016wsq,Smirnov:2019qkx,Klappert:2019emp,Peraro:2019svx,Klappert:2020aqs,DeLaurentis:2022otd,Magerya:2022hvj,Belitsky:2023qho,Liu:2023cgs,Mangan:2023eeb,Chawdhry:2023yyx,Maier:2024djk,Smirnov:2024onl}, special integral representations~\cite{Liu:2018dmc,Wang:2019mnn,Guan:2019bcx,Guan:2024byi}, and other methods~\cite{Barakat:2022qlc,Guan:2023avw,Chestnov:2024mnw} as well as machine learning techniques~\cite{vonHippel:2025okr,Song:2025pwy,Zeng:2025xbh}.

In this article we describe the new version \code{3} of \kira{} which is widely used in the community and backend in several public tools, see e.g.\ \citeres{Liu:2022chg,Magerya:2022hvj,Shtabovenko:2023idz,Chen:2024xwt,Prisco:2025wqs,Wu:2025aeg}.
After a brief recapitulation of \ac{ibp} reductions and an overview of the strategy of \kira{} in \cref{sect:prelimineries}, we discuss the new features in \cref{sect:main_features}.
The main features are improved seeding and selection strategies which reduce the reduction complexity significantly, especially if high tensor ranks are involved.
Furthermore, we discuss other new features like the possibility to add additional relations, sample the system over user-provided finite field points, perform symbolic reductions, and check the master integral basis.
In \cref{sect:changes} we discuss smaller changes with respect to the default behavior before quantifying the improvements with some benchmarks in \cref{sect:benchmarks} and concluding in \cref{sect:conclusions}.

\section{Preliminaries}
\label{sect:prelimineries}

The central object of this paper is the Feynman integral
\begin{align}
\label{eq:integralfamilydef}
  I_{\nu_{1}, \ldots, \nu_{n}}\left(\{s_i, m_i\}, d\right)=\int \left(\prod_{j=1}^{L} \mathrm{d}^{d} k_{j}\right)\,\prod_{j=1}^{N} \frac{1}{D_j^{\nu_{j}}} ,
\end{align}
where we follow the standard notations used in \citere{Maierhofer:2017gsa}.
We use the convention $D_j = q_{j}^{2}-m_{j}^{2}+\mathrm{i}\delta$, where $q_j$ denotes the momentum of the $j$-th propagator. The momentum $q_j$ can be written as a linear combination of $L$ loop momenta $k_i$ and $E$ linearly independent external momenta $p_j$ such that
\begin{align}
  q_j = \sum_{n=1}^L a_{jn} k_n + \sum_{n=1}^E b_{jn} p_n
\end{align}
for some integers $a_{jn}$ and $b_{jn}$. The term $\mathrm{i}\delta$ is the Feynman prescription.
Here, $\{s_i, m_i\}$ schematically denotes the set of external scales and internal masses on which the integral family depends. We work in dimensional regularization $d = d_{\mathrm{int}} - 2\varepsilon$, where $d_{\mathrm{int}}$ is a positive (even) integer. In the remainder of this paper, we will typically leave the dependence on the external scales, masses, and the dimension implicit. The propagator exponents $\nu_i$ are assumed to be integers.
The set of inverse propagators must be complete and independent in the sense that every scalar product of momenta can be uniquely expressed as a linear combination of $q_j$, squared masses $m_j^2$, and external kinematical invariants.
The number of propagators is thus $N=\frac{L}{2}(L+2 E+1)$ including auxiliary propagators that only appear with $\nu_j\le 0$.

The Feynman integrals of \cref{eq:integralfamilydef} are in general not independent.
In \citeres{Tkachov:1981wb,Chetyrkin:1981qh} it was found that they are related by so-called \acf{ibp} identities
\begin{equation}
  \int \left(\prod\limits_{i=1}^L \mathrm{d}^d k_i\right) \frac{\partial}{\partial k^\mu_i} \frac{P^\mu_j}{D_1^{\nu_1} D_2^{\nu_2} \cdots D_N^{\nu_N}} = 0 ,
\end{equation}
where $P^\mu_j$ is either a loop momentum, an external momentum, or a linear combination thereof.
In addition to these \ac{ibp} identities, there also exist so-called Lorentz-invariance identities~\cite{Gehrmann:1999as}
\begin{equation}
  \sum_{i=1}^E \left( p_i^\nu \frac{\partial}{\partial p_{i\mu}} - p_i^\mu \frac{\partial}{\partial p_{i\nu}} \right) I_{\nu_{1}, \ldots, \nu_{n}} = 0 .
\end{equation}
They do not provide additional information, but it was found that they often facilitate the reduction process.
Finally, there are symmetry relations between Feynman integrals which are most often searched for by comparing graph polynomials with Pak's algorithm~\cite{Pak:2011xt}.
All these relations provide linear equations of the form
\begin{equation}
  0 = \sum_i c_i\left(\{s_i, m_i\}, d\right) I_{\nu_{1}, \ldots, \nu_{n}}
  \label{eq:linear-relation}
\end{equation}
between different Feynman integrals, where the coefficients $c_i\left(\{s_i, m_i\}, d\right)$ are polynomials in the kinematic invariants $s_j$, the propagator masses $m_k$, and $d$.
The relations can be solved to express the Feynman integrals in terms of a basis of master integrals.
It was even shown that the number of master integrals for the standard Feynman integrals of \cref{eq:integralfamilydef} is finite~\cite{Smirnov:2010hn}.

Nowadays, the most prominent solution strategy is the Laporta algorithm~\cite{Laporta:2000dsw}:
The linear relations of \cref{eq:linear-relation} are generated for different values of the propagator powers $\nu_j$ resulting in a system of equations between specific Feynman integrals.
Each choice of $\{\nu_j\}$ is called a \emph{seed} and this generation process is also known as \emph{seeding}.
The system can then be solved with Gauss-type elimination algorithms.

This requires an ordering of the integrals.
In \kira{} we first assign the integrals to so-called topologies based on their respective sets of propagators, i.e.\ their momenta and masses.
We assign a unique integer number to each topology.
Secondly, we assign each integral to a sector
\begin{equation}
  S = \sum\limits_{j=1}^N 2^{j-1}\,\theta(\nu_j - \tfrac{1}{2}),
  \label{eq:sector}
\end{equation}
where $\theta(x)$ is the Heaviside step function.
A sector $S$ (with propagators powers $\nu_j$) is called subsector of another sector $S'$ (with propagators powers $\nu'_j$) if $S<S'$
and $\theta(\nu_j - \tfrac{1}{2})\le \theta(\nu'_j - \tfrac{1}{2})$ for all $j=1,\dots,N$.
We denote as top-level sectors those sectors which are not subsectors of other sectors that contain Feynman integrals occurring in the reduction problem at hand.
For the discussions in the following sections it is beneficial to use the big-endian binary notation instead of the sector number defined by \cref{eq:sector}, i.e.\ the sector $S=5$ is represented by b$1010$ where each number represents one bit of the number $5$ in reverse order.
In this notation it is immediately visible that b$1010$ is a subsector of b$1110$.

Furthermore, it is useful to define the number of propagators with positive powers,
\begin{equation}
  t = \sum_{j=1}^N \theta(\nu_j - \tfrac{1}{2}),
  \label{eq:def-t}
\end{equation}
and
\begin{equation}
  r = \sum_{j=1}^N \nu_j \theta(\nu_j - \tfrac{1}{2}),\qquad
  s = -\sum_{j=1}^N \nu_j\theta(\tfrac{1}{2} - \nu_j),\qquad
  d = \sum_{j=1}^N (\nu_j - 1) \theta(\nu_j - \tfrac{1}{2})
  \label{eq:rsd}
\end{equation}
denoting the sum of all positive powers, the negative sum of all negative powers, and the sum of positive powers larger than 1 (referred to as \emph{dots}), respectively.

On the one hand, these concepts allow us to order the integrals in \kira{}.
On the other hand, they are used to limit the seeds $\{\nu_j\}$ for which equations are generated.
We generate the equations only for those sets $\{\nu_j\}$ for which $r\le r_{\mathrm{max}}$, $s\le s_{\mathrm{max}}$, and $d\le d_{\mathrm{max}}$, where $r_{\mathrm{max}}$, $s_{\mathrm{max}}$, and $d_{\mathrm{max}}$ are input provided by the user.

The workflow in \kira{} is split into two components: first generating the system of equations and then solving it.
We offer two different strategies for the latter.
First, the system can be solved analytically using the computer algebra system \texttt{Fermat}~\cite{Fermat} for the rational function arithmetic.
Secondly, the system can be solved using modular arithmetic over finite fields~\cite{Kauers:2008zz,Kant:2013vta,vonManteuffel:2014ixa,Peraro:2016wsq} with our internal solver \pyred{}.
The variables, i.e.\ kinematic invariants and masses, are replaced by random integer numbers and all arithmetic operations are performed over prime fields.
Choosing the largest $63$-bit primes allows us to perform all operations on native data types and to avoid large intermediate expressions.
The final result can then be interpolated and reconstructed by \emph{probing} the system sufficiently many times for different random choices of the variables.
We employ the library \texttt{FireFly} for this task~\cite{Klappert:2019emp,Klappert:2020aqs}.

The main improvements of \kirathr{} concern the generation of the system of equations and it is thus beneficial for the understanding of the reader to summarize the old algorithm from version \code{2.3} in more detail:
\begin{enumerate}
  \item Identify trivial sectors and discard all integrals in these sectors.
  \item Identify symmetries between sectors including those of other integral families and collect them in the set $\{S'\}$.
  \item Generate \ac{ibp} equations for all sectors not in $\{S'\}$ and generate only symmetry equations for all sectors in $\{S'\}$.
  Start from the simplest one and generate all equations within the specifications provided by the user:
  \begin{enumerate}[label=\roman*)]
    \item Generate one equation. Check if it is linearly independent by inserting all previously generated equations numerically~\cite{Kant:2013vta}.
    \item Keep, if it is linearly independent, otherwise drop it.
  \end{enumerate}
  \item After all equations were generated, solve the system for the target integrals specified by the user with modular arithmetic and drop all equations which do not contribute to their solutions.
\end{enumerate}
At the end of this procedure \kira{} has generated a system of linearly independent equations specifically trimmed to suffice to reduce the target integrals to master integrals.

\section{New features}
\label{sect:main_features}

\subsection{Improved seeding}
\label{ssec:seeding}

The choice of which equations should be generated for the reduction has severe impact on the performance of the reduction.
Previous versions of \kira{} seed conservatively and generate all equations within the bounds specified by the user to reduce to the minimal basis of master integrals.
Irrelevant equations are filtered out through a selection procedure.
However, for integrals with many loops or high values of $r$, $s$, and $d$ the combinatorics of the seeds easily gets out of hand and already makes the generation of the system of equations unfeasible, running both into runtime and memory limits.

We show that it is possible to find a subset of relevant equations which improves the generation of equations lifting the limits for many applications, as demonstrated in \citeres{Driesse:2024xad,Driesse:2024feo}.
We revise the seeding process in \kira{} and identify three areas for improvement: sectors related by symmetries, sectors containing preferred master integrals, and subsectors.
We discuss the first two points in \cref{sssec:seeding-changed} and the latter in \cref{sssec:truncate}.

\subsubsection{Changed behavior for sectors with symmetries or preferred master integrals}
\label{sssec:seeding-changed}

In previous versions of \kira{}, sectors which are related by symmetries to other sectors are seeded with the same values for $r_{\mathrm{max}}$ and $s_{\mathrm{max}}$, ignoring any set $d_{\mathrm{max}}$.
Similarly, sectors which contain one of the preferred master integrals are seeded with the maximum values of $r_{\mathrm{max}}$ and $s_{\mathrm{max}}$ for all sectors, again ignoring $d_{\mathrm{max}}$.
Those choices turn out to be too conservative and negatively impact the performance of \ac{ibp} reductions.
With \kirathr{} subsectors inherit the values for $r_{\mathrm{max}}$, $s_{\mathrm{max}}$, and also $d_{\mathrm{max}}$ from higher sectors, independently of symmetries and preferred master integrals.

Furthermore, in previous versions \kira{} did not generate \ac{ibp} equations for sectors which can be mapped away by symmetries and instead relied on symmetry relations alone.
While this is a good strategy to generate fewer equations, symmetry equations quickly become longer with more complicated coefficients when they are generated with numerators.
Their complexity easily overtakes those of \ac{ibp} equations even for moderate values of $s$ and makes the solution significantly more expensive.
Hence, with \kirathr{} we also generate \ac{ibp} equations for sectors which are mapped away by symmetries.
This increases the reduction performance, while the growth of generated equations is easily compensated by the new truncation option discussed in the next subsubsection.

\subsubsection{Truncating the seeds}
\label{sssec:truncate}

By default, the values for $r_{\mathrm{max}}$ and $s_{\mathrm{max}}$ for one sector are inherited by all subsectors.
For example, if we are interested in the integral $T(1,1,1,1,1,1,1,0,-4)$, the typical choice is to seed the sector b$111111100$ with $r_{\mathrm{max}} = 7$ and $s_{\mathrm{max}} = 4$.
This then propagates to the lower sectors and sector b$111111000$ is automatically seeded with $r_{\mathrm{max}} = 7$, corresponding to $d_{\mathrm{max}} = 1$, and $s_{\mathrm{max}} = 4$ and sector b$111000000$ with $r_{\mathrm{max}} = 7$, corresponding to $d_{\mathrm{max}} = 4$, and $s_{\mathrm{max}} = 4$, respectively.
The increase in $d_{\mathrm{max}}$ can already be restricted by manually setting $d_{\mathrm{max}}$ (with the exceptions discussed in \cref{sssec:seeding-changed}), but we previously argued against this to prevent finding too many master integrals.
With the new release of \kirathr{} we revise this recommendation and encourage setting $d_{\mathrm{max}}$.

However, keeping $s_{\mathrm{max}}$ constant when descending to the subsectors is even more problematic because the combinatorics of distributing irreducible scalar products in lower sectors grows factorially.
As it turns out, most of these seeds are actually not required to reduce integrals with positive $s$ in higher sectors and one can decrease $s_{\mathrm{max}}$ in lower sectors.
In \kirathr{} we introduce a new feature to restrict $s$ following the function
\begin{equation}
  s \leq \max(1,\ t - l + 1,\ s_{\text{max},\text{sector}})
  \label{eq:truncation}
\end{equation}
where $t$ was defined in \cref{eq:def-t}, $l$ can be chosen by the user, and $s_{\text{max},\text{sector}}$ is the value that was explicitly set by the user for this sector with the option \code{reduce}.
If no explicit value is provided with \code{reduce}, $s_{\text{max},\text{sector}} = 0$, i.e.\ it is not inherited from higher sectors.
The usefulness of this feature was first described and proven in \citere{Driesse:2024xad}.
Similar observations were also made by the developers of \code{Blade}~\cite{Guan:2024byi} and \code{FIRE}~\cite{Bern:2024adl}.

This feature can be enabled with the new job file option
\begin{verbatim}
    truncate_sp:
      - {topologies: [<topologies>], l: <n>}
\end{verbatim}
with the two arguments
\begin{itemize}
  \item \code{topologies} (optional): a list of topologies to which \code{truncate\_sp} should be applied.
  If no topology is provided, \kira{} applies \code{truncate\_sp} to all topologies.
  \item \code{l} (mandatory):
  This option takes the integer argument \code{<n>} (negative numbers are allowed) and instructs \kira{} to choose seeds following \cref{eq:truncation} with $l = \code{<n>}$.
\end{itemize}
A good choice of \code{<n>} can result in a speedup of orders of magnitude.
To choose the optimal parameters we recommend the following strategy:
\begin{enumerate}
    \item The highest possible value for \code{<n>} is computed by taking the difference between the $t$ value of the top-level sector and the highest number $s$ of an integral which one wants to reduce.
    This choice is too optimistic in almost all cases and one should run \kira{} to check if all integrals reduce to the expected master integrals.
    The new option \code{check\_masters} described in \cref{ssec:check-masters} is helpful for this purpose.
    If too many master integrals are reported, we recommend to tune the options provided to \kira{} starting with those sectors with the highest value of $t$ that report too many master integrals.
    \item\label{enu:tune-2} Tune all sectors at level $t$:
    \begin{enumerate}[label=\roman*)]
      \item If many or even all sectors with the same value of $t$ have too many master integrals whose values are cut off by \cref{eq:truncation}, decrease \code{<n>} to a value that seeds those additional master integrals.
      \item If only a few sectors have too many master integrals cut off by \cref{eq:truncation} or $s > s_{\mathrm{max}}$, increase $s_{\mathrm{max}}$ for the sector.
      \item If there are too many master integrals with $d > d_{\mathrm{max}}$ in a sector, increase $d_{\mathrm{max}}$ for this sector.
    \end{enumerate}
    \item\label{enu:tune-3} Run \kira{} again to check if there are additional master integrals on level $t$.
    If yes, tune those sectors further by repeating steps \ref{enu:tune-2} and \ref{enu:tune-3}.
    If not, reduce $t$ by one and repeat steps \ref{enu:tune-2} and \ref{enu:tune-3} on this level.
\end{enumerate}

Using the option \texttt{select\_mandatory\_recursively} is not recommended if \texttt{truncate\_sp} is enabled because the aforementioned option selects all integrals to reduce with a constant value of $s_{\mathrm{max}}$.
This results in \kira{} reporting many unreduced integrals because many required seeds are removed by \texttt{truncate\_sp}.

Summarizing this section, our recommendation for the job file with \kirathr{} reads
\begin{verbatim}
jobs:
 - reduce_sectors:
    reduce:
      - {topologies: [topo7], sectors: [b111111100], r: 7, s: 5, d: 0}
      - {topologies: [topo7], sectors: [b011100100], r: 5, s: 3, d: 1}
    truncate_sp:
      - {topologies: [topo7], l: 3}
    select_integrals:
      select_mandatory_list:
        - [seeds7]
\end{verbatim}
for the example topology \code{topo7} provided with the source code.
The top-level sector \code{b111111100} (\code{127}) is seeded with $r_{\mathrm{max}} = 7$, $s_{\mathrm{max}} = 5$, and $d_{\mathrm{max}} = 0$, providing sufficient seeds to reduce all integrals in this sector.
Furthermore, \code{truncate\_sp} cuts away seeds in subsectors.
However, sector \code{b011100100} (\code{78}) had to be tuned by increasing $d_{\mathrm{max}}$ to $1$.

\subsection{Improved selection algorithm}
\label{ssec:selection}

\begin{figure}[ht]
  \centering
  \includegraphics[scale = 0.7]{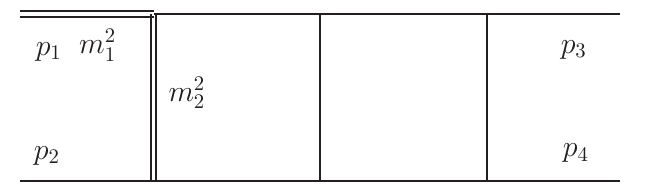}
  \caption{The planar double box \texttt{topo7} contributing, e.g., to single-top production.}
  \label{fig:topo7}
\end{figure}

Once \kira{} determines the linearly independent system of equations, it will select the relevant equations to reduce specific integrals as defined for example with the option \texttt{select\_integals}.
The algorithm to select relevant equations in previous \kira{} versions is not optimal.
One weakness can easily be illustrated when reducing all integrals in the top-level sector \code{b111111100} (\code{127}) of the topology \texttt{topo7} depicted in \cref{fig:topo7} with exactly $s=5$.
With \kiratwothr{}, we first seed with $s_{\mathrm{max}}=5$ and, secondly, with $s_{\mathrm{max}}=6$, and in both cases select equations to reduce the same set of integrals.
As can be seen in \cref{tab:selection}, the number of selected equations and with it the number of terms as well as the time to solve the system increases.
\begin{table}[ht]
  \begin{center}
    \caption{Comparison of the generated and selected systems for \texttt{topo7} when seeded with different values of $s_{\mathrm{max}}$.}
    \label{tab:selection}
    \begin{tabular}{c|c|c|c|c}
      \toprule
      & \texttt{2.3}, $s_{\mathrm{max}}=5$ & \texttt{2.3}, $s_{\mathrm{max}}=6$ & \texttt{3.0}, $s_{\mathrm{max}}=5$ & \texttt{3.0}, $s_{\mathrm{max}}=6$ \\
      \midrule
      \# of generated eqs. & 3\,317\,357 & 6\,009\,193 & 4\,053\,617 & 7\,511\,785 \\
      \# of selected eqs. & 62\,514 & 85\,119 & 30\,984 & 30\,984 \\
      \# of terms & 461\,336 & 659\,803 & 245\,975 & 245\,975 \\
      \bottomrule
    \end{tabular}
  \end{center}
\end{table}
This can be explained by the fact that \kira{} generates and selects equations with the sector-by-sector approach as discussed in \cref{sect:prelimineries}.
It prefers equations generated with higher values for $r$, $s$, and $d$ in lower sectors, which are usually more complicated than equations generated with low values of $r$, $s$, and $d$ in higher sectors.

Secondly, we identified another problem: \kira{} selects equations which do not contribute to the final result because they cancel in intermediate steps.
This can be illustrated with the toy system
\begin{equation}
  \begin{aligned}
    (\mathrm{i}) \quad 0=&\,x[6]+b_1\,x[2] , \\
    (\mathrm{ii}) \quad 0=&\,x[6]+c_1\,x[1] , \\
    (\mathrm{iii}) \quad 0=&\,x[5]+a_1\,x[2]+a_2\,x[4]+a_3 x[3] , \\
    (\mathrm{iv}) \quad 0=&\,x[5]+a_1\,x[2] ,
  \end{aligned}
  \label{eq:select-toy-initial}
\end{equation}
where $x[i]$ represents a Feynman integral and the Roman number in front denotes the equation number.
We now solve this system using Gaussian elimination.

In \kira{} we solve equations one at a time, starting with equation $(\mathrm{iv})$ in \cref{eq:select-toy-initial} and work in a bottom-up approach.
Whenever we add a new equation, \kira{} immediately inserts already known solutions into the new equation.
If the equation equates to zero, it is treated as redundant and is discarded.
Otherwise its terms are sorted with respect to some Laporta ordering and the new equation is added to the solved equations.
This is repeated until all equations are processed.
This step is called forward elimination in \kira{} and results in a system that is in upper-triangular form.
After the forward elimination \cref{eq:select-toy-initial} reads
\begin{equation}
  \begin{aligned}
    (\mathrm{i}) \quad 0=&\,b_1\,x[2]-c_1\,x[1] \quad &\{\mathrm{ii}\}& , \\
    (\mathrm{ii}) \quad 0=&\,x[6]+c_1\,x[1] \quad &\{\}& , \\
    (\mathrm{iii}) \quad 0=&\,a_2\,x[4]+a_3\,x[3] \quad &\{\mathrm{iv}\}& , \\
    (\mathrm{iv}) \quad 0=&\,x[5]+a_1\,x[2] \quad &\{\}&.
  \end{aligned}
  \label{eq:select-toy-forward}
\end{equation}
\kira{} traces which equations were used to eliminate integrals from the newly added equation, in this example denoted by the Roman numbers in the curly brackets:
Equation $(\mathrm{ii})$ was used to eliminate the integral $x[6]$ in equation $(\mathrm{i})$ and equation $(\mathrm{iv})$ was used to eliminate the integral $x[5]$ in equation $(\mathrm{iii})$.

In the second step, the backward substitution iterates the system of equations bottom-up and inserts all available solutions.
Again we record each equation used to eliminate integrals.
After the back substitution we obtain
\begin{equation}
  \begin{aligned}
    (\mathrm{i}) \quad 0=&\,b_1\,x[2]-c_1\,x[1] \quad &\{\mathrm{ii}\}& , \\
    (\mathrm{ii}) \quad 0=&\,x[6]+c_1\,x[1]\, \quad &\{\}& , \\
    (\mathrm{iii}) \quad 0=&\,a_2\,x[4]+a_3\,x[3] \quad &\{\mathrm{iv}\}& , \\
    (\mathrm{iv}) \quad 0=&\,x[5]+a_1\,c_1/b_1\,x[1] \quad &\{\mathrm{i}, \mathrm{ii}\}& .
  \end{aligned}
  \label{eq:select-toy-final}
\end{equation}
Here equation $(\mathrm{i})$ is used to eliminate the integral $x[2]$ in equation $(\mathrm{iv})$.

In previous versions the information in the curly brackets from \cref{eq:select-toy-final} is used to select relevant equations for the reduction process.
For the reduction of the integral $x[4]$, the previous versions of \kira{} selected equation $(\mathrm{iii})$ because it provides the solution for $x[4]$, then equation $(\mathrm{iv})$ which was inserted into equation $(\mathrm{iii})$, and finally equations $(\mathrm{i})$ and $(\mathrm{ii})$ which were inserted into equation $(\mathrm{iv})$.
However, the latter two do not provide any relevant information: it is easy to spot in \cref{eq:select-toy-initial} that equation $(\mathrm{iii})$ contains equation $(\mathrm{iv})$ as a subequation.
This subequation thus vanishes after inserting equation $(\mathrm{iv})$ only and the information of equations $(\mathrm{i})$ and $(\mathrm{ii})$ is not needed.
We call this phenomenon \emph{hidden zero} which in general can take more complicated forms.

To address both the deficits of the sector-by-sector approach and the hidden zeros, \kirathr{} selects the relevant equations after the forward elimination only.
For the example above we can clearly see that this information in \cref{eq:select-toy-forward} is sufficient to reduce the integral $x[4]$ to master integrals and it gives the relevant equations only.
For more generic examples the algorithm requires one additional step:
\kira{} reduces the system numerically over a finite field once and checks if the selected equations are sufficient to reduce all integrals to the minimal basis of master integrals.
All unreduced integrals are added to the list and the reduction is repeated until all requested integrals are reduced to the minimal basis.
This algorithm is generic and applies to user-defined systems as well.

However, it cannot give the minimal number of equations for general problems as more stealthy zeros in the forward elimination remain undetected.
One could heuristically reduce the number of equations by randomly reordering the equations and repeating the algorithm discussed above, throwing out all irrelevant equations at the end of every iteration.
Repeating this process, we end up with less and less equations until reaching a local minimum.
However, a randomly reordered system of equations typically gets dense after a few Gaussian elimination steps, making the solution computationally expensive.
One could reorder the system of equations in bunches, e.g.\ in batches of 20 equations, but in our experiments this yield at most $20$\,\% less equations.
This is a very low improvement from our perspective and does not justify the implementation of this algorithm yet.
There is another algorithm, privately called the \textit{Oerlikon algorithm}, designed to select necessary \ac{ibp} equations, see \citeres{Bohm:2018bdy,Wu:2023upw}, which we plan to investigate with \kira{} in the future.

The impact of the new selection algorithm is illustrated in \cref{tab:selection}.
Since the new truncation discussed in \cref{sssec:truncate} option is not enabled, \kirathr{} generates more equations than \kiratwothr{} as discussed in \cref{sssec:seeding-changed}.
However, even though $22$\,\% more equations are generated for $s_{\mathrm{max}} = 5$, the new selection algorithm actually selects $50$\,\% less equations, indicating that many hidden zeros and irrelevant equations are dropped.
Moreover, increasing $s_{\mathrm{max}}$ to $6$ does not change the selected equations anymore as in previous versions.

\subsection{Extra relations}

There can be additional relations between integrals that are not discovered by \kira{}, usually through some non-trivial symmetries, see e.g.\ \citeres{Abreu:2022vei,Buccioni:2023okz,Wu:2024paw}, or four-dimensional kinematic constraints, see e.g.\ \citere{Abreu:2024fei}.
Therefore, it is desirable to add those relations to the system generated by \kira{}.
In previous version additional relations could only be added in the user-defined system mode.
With \kirathr{} it is now possible to add them during the normal reduction setup with the option \code{extra\_relations:\ <file>} in the job file.
It takes exactly only one file as argument in which the relations should be provided in the format of user-defined systems as described in \citere{Klappert:2020nbg}.
The additional equations are inserted after the \ac{ibp} relations for a specific sector were generated.
Additional relations belonging to sectors beyond the seeded sectors are inserted at the end of the generation process.

\subsection{Numerical sampling}
\label{subsec:numerical_points}

For complicated reductions it might be useful to sample the reduction numerically on chosen phase space points, either before attempting to obtain full analytic expressions or bypassing them altogether, see e.g.\ \citere{Agarwal:2024jyq}.
In the context of \kira{} this feature has been explored in \citere{Abreu:2024fei}.

There are two aspects with respect to this feature:
First, we implemented an interface for the user to provide both the prime for the finite field and the sample points with the option \code{numerical\_points:\ <file>} where the file format reads
\begin{verbatim}
prime 9223372036854771977
d s t m12 m22
11111 -1 -38 -1039/6 -2712776
11112 -1 -38 -1039/6 -2712776

prime 9223372036854771689
d s t m12 m22
11111 -1 -38 -1039/6 -2712776
11112 -1 -38 -1039/6 -2712776
\end{verbatim}
The first line specifies the user-defined prime number for a finite field.
The largest prime number currently permitted is $9223372036854775783$, i.e.\ the largest $63$-bit prime.
The second line lists all symbols that the user intends to assign numerical values to.
It should contain all symbols appearing in the system, i.e.\ all kinematic invariants, the dimension $d$, and potentially further symbols in user-defined systems of equations.
In the following lines the numerical values corresponding to those symbols are provided.
An empty line separates the collection of points from another collection of points in a different prime field.

The second, less obvious aspect concerns the generation of the system of equations by \kira{}.
By default \kira{} chooses pseudo-random numbers for all symbols in the system during the generation step to check the equations for linear dependence.
However, the user might restrict the sample points to a slice in phase space in which not all of the symbols are independent.
In this case, further cancellations can occur and a system of linearly-independent equations generated on a general pseudo-random phase space point might no longer contain sufficient information to reduce all integrals.
Therefore, when instructed with \code{numerical\_points} in the job file, \kira{} generates the system with the first provided phase space point and prime and writes it to disk as usual.
The same result can be obtained with the command-line options \code{----set\_value} and \code{----prime\_user}, see also \cref{ssec:set-value}.
By default, \kira{} writes the system to disk with algebraic coefficients, but can be instructed to replace all symbols by the given values with the job file option \code{symbols2num:\ true}.

The results of the numerical reductions are written to the files \code{results/<topology>/<input file name>\_<prime>\_<j>.m}.
Since we want to avoid to produce too large files, only 1000 entries are written into each file.
If this limit is reached, \kira{} writes to the next file with \code{j} increased by one.
An example output for a reduction with the example integral family \code{topo7} looks like
\begin{verbatim}
{9223372036854771977,
  {{11111, 9223372036854771976, 9223372036854771939, 7686143364045643141,
  9223372036852059201},
    {
      {topo7[1,1,1,1,1,1,1,-5,0],
        {topo7[1,1,1,1,1,1,1,0,-1], 2951302433002828169},
        {topo7[1,1,1,1,1,1,1,-1,0], 819004719938417371},
        {topo7[1,1,1,1,1,1,1,0,0], 5495110817388584284},
        ...
      },
      ...
    }
  },
  {{11112, 9223372036854771976, 9223372036854771939, 7686143364045643141,
  9223372036852059201},
    {
      {...}
    }
  }
}
\end{verbatim}
where the indentation is for better illustration.
Each file is structured as a nested list in \code{Mathematica} readable format.
The first entry is a prime number used in the computations, followed by sublists, each containing \ac{ibp} results for a specific numerical point.
Each sublist starts with the numerical point as an identifier (note that the user input has been converted into a finite field element over the user provided prime), followed by the numerical reduction results.
Each reduction result is a sublist beginning with the integral to be reduced, followed by pairs of master integrals and their corresponding reduction coefficients.

\subsection{Symbolic integration-by-parts reductions}
\label{subsec:symbolic_ibp}

Obtaining general recurrence relations for arbitrary propagator powers can be seen as the ultimate goal for \ac{ibp} reductions since they should facilitate an efficient reduction for all integrals of a family.
\code{LiteRed}~\cite{Lee:2012cn,Lee:2013mka} is the only public tool that offers an algorithm to search for such relations, but it is not guaranteed to succeed.
Recently, a different strategy has been explored in \citere{Guan:2023avw}.

In \kirathr{} it is now possible to treat propagator powers symbolically and use the Laporta strategy to derive symbolic recursion relations.
Every power is treated as an additional symbol which makes it computationally expensive to derive fully algebraic recursion relations in all indices.
However, constructing recursion relations for one or very few powers works reasonably well in many problems.
This feature is activated with the option \code{symbolic\_ibp:\ [<n1>, <n2>, ...]} in \code{integralfamilies.yaml}.
\code{<n1>}, \code{<n2>}, \dots may take an integer number between $1$ and the total number of propagators provided with the integral family definition.
The symbolic power is then associated with the \code{n1}, \code{n2}, \dots propagator from top to bottom as listed in \code{integralfamilies.yaml}.

In the following we illustrate our best strategy for a symbolic \ac{ibp} reduction as applied in \citere{Fael:2023tcv} with an example available in the directory \code{examples/symbolic\_IBP} provided with the source code.
It is advised to take as a top-level sector all propagators which have positive propagator powers, but to omit the propagators associated with the symbolic powers.
The relevant part of \code{integralfamilies.yaml} may read
\begin{verbatim}
top_level_sectors: [b011111100] # b011111100 is same as sector 126
symbolic_ibp: [1]
\end{verbatim}
where we treat the first propagator power as a symbolic power.
The example job file \code{jobs\_lowering.yaml} is used to generate lowering relations for the symbolic power:
\begin{verbatim}
jobs:
 - reduce_sectors:
    reduce:
      - {topologies: [topo7], sectors: [b011111100], r: 7, s: 2, d: 1}
    truncate_sp:
      - {topologies: [topo7], l: 4}
    select_integrals:
      select_mandatory_list:
        - [toReduce_lowering]
    preferred_masters: masters_lowering
    integral_ordering: 2
    run_initiate: true
    run_firefly: true
 - kira2math:
    target:
     - [toReduce_lowering]
\end{verbatim}
The file \texttt{toReduce\_lowering} contains one integral which will be reduced with \kira{}:
\begin{verbatim}
#to reduce
topo7[0,1,1,1,1,1,1,-2,0]
\end{verbatim}
The essential master integrals are listed in the file \texttt{masters\_lowering}:
\begin{verbatim}
topo7[-1,1,1,1,1,1,1,0,0]
topo7[-1,1,1,1,1,1,1,0,-1]
topo7[-1,1,1,1,1,1,1,-1,0]
topo7[-2,1,1,1,1,1,1,0,0]
\end{verbatim}
Since the first index is symbolic, the corresponding entries are interpreted as lowering or raising operations.
The first indices of the selected integrals serve as starting point and, if all of them are larger than the first indices of all preferred master integrals, \kira{} will construct lowering relations, and raising operations if the first indices of the selected integrals are smaller.
In the example above \kira{} will try to construct a lowering relation shifting the first index down by $-1$ and $-2$.
If all of the selected integrals and all of the preferred master integrals have zero or negative symbolic indices, the corresponding sector is merged with a subsector, generating less equations.
In the example above sectors \code{b011111100} (\code{126}) and \code{b111111100} (\code{127}) are considered to be the same.

In the resulting expressions, the symbols \code{b0}, \dots,\code{b(N-1)} denote symbolic powers, where \code{N} is the total number of propagators listed in \code{integralfamilies.yaml}.
In the example above, we encounter the relation
\begin{verbatim}
topo7[0,0,1,0,0,0,1,-1,0] ->
 + topo7[-1,0,1,0,0,0,1,0,0]*(((d+(-2))*(1))/((b0+(-1))*(1)))
\end{verbatim}
in the results directory, which can be translated into the recursion relation
\begin{equation}
  \code{topo7}(b_0,0,1,0,0,0,1,-1,0) \rightarrow \frac{d-2}{b_0-1} \code{topo7}(b_0-1,0,1,0,0,0,1,0,0)
 \end{equation}
 in the first $b_0$ index.

\subsection{Check master integral basis}
\label{ssec:check-masters}

In many cases users determine and choose the basis of master integrals before performing a full-fledged reduction.
Criteria might be a convenient basis choice for the reduction and/or for solving the master integrals.
Since version \code{1.1}, \kira{} allows users to provide their basis choice with the job file option \texttt{preferred\_masters}.
Since version \code{2.0}, the basis can also consist of linear combinations of master integrals as described in \citere{Klappert:2020nbg}.
However, by default \kira{} supplements the provided basis with additional master integrals in case it cannot reduce to the provided basis.
With the new job file option \texttt{check\_masters:\ true}, \kirathr{} compares the master integrals it found when generating the system with those provided by \texttt{preferred\_masters}.
It reports both additional master integrals as well as those provided master integrals it does not need for the reduction.
In case it finds additional master integrals, it aborts the reduction and throws an error.

This feature is especially useful if users already determined the basis and try to exploit the new seeding features discussed in \cref{ssec:seeding} to generate an optimal system of equations.

\section{Further changes}
\label{sect:changes}

\subsection{Internal reordering of propagators}
\label{ssec:order}

\begin{figure}[ht]
  \centering
  \includegraphics[trim = 36 538 36 36, scale = 0.4]{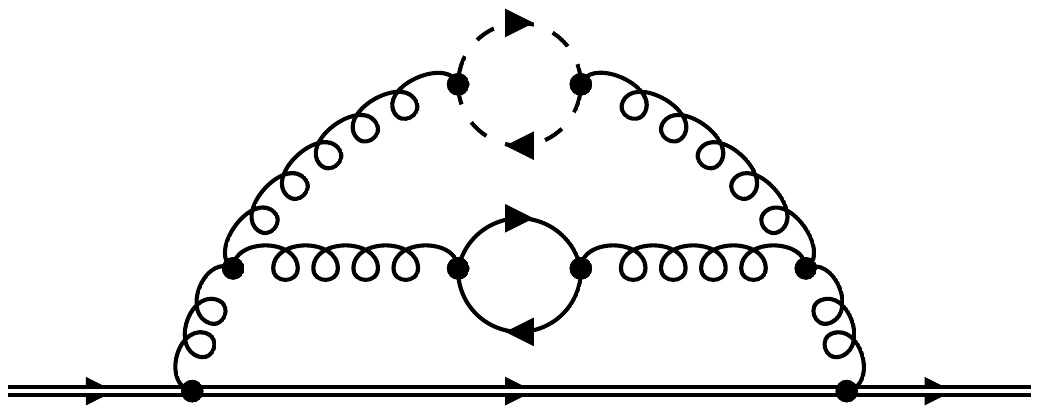}
  \caption{Four-loop propagator topology with two masses.}
  \label{fig:four-loop-propagator}
\end{figure}

It is well known that the definition of the propagators in \cref{eq:integralfamilydef} influences the performance of the reduction.
Not only different choices for the momenta have an impact, simply reordering them can make a difference, especially when selecting equations before the actual reduction as in \kira{}.
This can be demonstrated with the four-loop propagator topology depicted in \cref{fig:four-loop-propagator}.
We reduce all integrals appearing in the quark self energy amplitude for the relation between the quark mass renormalization constants in the $\overline{\text{MS}}$ and the on-shell schemes~\cite{Fael:2021kyg}.
We compare the propagator order
\begin{equation}
  \begin{split}
    &p_1^2 - m_1^2,\ (p_1 - q)^2,\ \underline{(p_2 + p_3)^2},\ (p_1 - p_3 + p_4 - q)^2,\ (p_3 - p_4)^2,\ \underline{(p_3 + q)^2},\ \underline{(p_2 + p_4)^2}, \\
    &p_2^2 - m_2^2,\ (-p_1 + p_2 + p_3 - p_4 + q)^2 - m_2^2,\ p_3^2, p_4^2,\ \underline{(p_1 + p_3)^2},\ \underline{(p_1 + p_4)^2},\ \underline{(p_1 + p_2)^2} ,
  \end{split}
  \label{eq:propagator-order-1}
\end{equation}
which was automatically chosen when generating the amplitude in \citere{Fael:2021kyg}, denoted as \textit{bad choice}, with propagator order 1,
\begin{equation}
  \begin{split}
    &p_3^2,\ p_4^2,\ p_1^2 - m_1^2,\ p_2^2 - m_2^2,\ (p_1 - q)^2,\ (p_3 - p_4)^2,\ (p_1 - p_3 + p_4 - q)^2,\\
    &(-p_1 + p_2 + p_3 - p_4 + q)^2 - m_2^2,\ \underline{(p_2 + p_3)^2},\ \underline{(p_3 + q)^2},\ \underline{(p_2 + p_4)^2},\ \underline{(p_1 + p_3)^2},\\
    &\underline{(p_1 + p_4)^2},\ \underline{(p_1 + p_2)^2} ,
  \end{split}
  \label{eq:propagator-order-2}
\end{equation}
which implements the guidelines outlined in \citere{Maierhofer:2018gpa}:
order by the number of momenta and put massless propagators first.
The underlined entries denote auxiliary propagators which only appear in the numerator and should be placed at the end.

When comparing the systems of equations generated with \kirathr{} for the two permutation ordering choices in \cref{tab:propagator-order}, we see that the system generated with order 1 selects 3\,\% less equations, 6.5\,\% less terms, and can be solved 29\,\% faster for a phase space point over a finite field than the system generated with the \textit{bad choice}.
\begin{table}[ht]
  \begin{center}
    \caption{Comparison of the generated and selected systems of equations for the two propagator orders defined in \cref{eq:propagator-order-1,eq:propagator-order-2} of the four-loop propagator topology depicted in \cref{fig:four-loop-propagator}.}
    \label{tab:propagator-order}
    \begin{tabular}{c|c|c}
      \toprule
      & bad choice & order 1 \\
      \midrule
      \# of equations & 481\,855 & 467\,338 \\
      \# of terms & 5\,902\,637 & 5\,537\,560 \\
      $T_{\pyred{}}$ & 8.0\,s & 6.2\,s \\
      \bottomrule
    \end{tabular}
  \end{center}
\end{table}

Since version \kiratwothr{} offers the possibility to reorder the propagators internally independent of the order specified in the integral family definition, i.e.\ the propagator order in the input and output files differs from the order \kira{} uses internally to perform the reduction.
The ordering can either be chosen manually or be adjusted automatically according to four different ordering schemes.
The one outlined above usually provides the best results.
With \kirathr{} we will automatically turn on this ordering scheme by default.
It can be overridden in \code{config/integralfamilies.yaml} by choosing a different ordering scheme with the option \code{permutation\_option:\ <N>}, with \code{<N>} taking values from 1 to 4, or by manually fixing the order with the option \code{permutation:\ [<n1>,\dots,<nN>]}, where each \code{<ni>} denotes the number of the i'th propagator defined for the topology.

\subsection{Support for 128-bit weights}

Internally \kira{} represents all integrals by an integer weight.
Up to \code{Kira\;2.2} they were limited to native 64-bit integers.
However, those do not suffice to represent all integrals for some new, ambitious projects, see e.g.\ \citere{Fael:2023tcv}.
Hence, with \kiratwothr{} we introduced 128-bit weights based on compiler extensions available in standard \code{C++} compilers like \code{GCC}~\cite{gcc} and \code{clang}~\cite{clang}.
The width of the weights can be set with
\begin{verbatim}
meson setup -Dweight_width=<WIDTH> build
\end{verbatim}
when compiling \kira{}, where \code{<WIDTH>} can either be \code{64} (default) or \code{128}.
If \code{128} is chosen, the executable is called \code{kira128}.

Note that sectors are still represented by a 32-bit integer and, thus, topologies with more than 31 lines are not yet supported.

\subsection{Command Line option \texttt{----set\_value}}
\label{ssec:set-value}

When generating a system of equations with previous versions of \kira{}, a pseudo-random point was selected to determine the linearly independent system.
The old implementation of \code{----set\_value} did not influence this point.
In the same way as the job file option \code{numerical\_points} introduced in \cref{subsec:numerical_points}, \kirathr{} now ensures that user-defined numerical points are respected during the generation and selection steps.
In combination with the job option \texttt{symbols2num:\ false} the resulting independent system of equations is written to disk with symbolic coefficients.

\section{Benchmarks}
\label{sect:benchmarks}

\begin{figure}[ht]
  \centering
  \begin{subfigure}[b]{0.5\textwidth}
    \centering
    \includegraphics[scale=0.7]{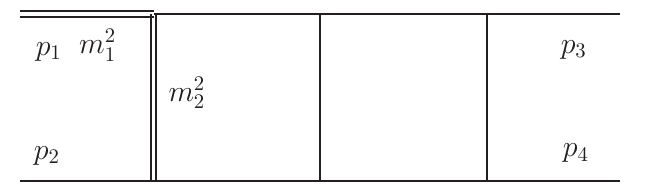}
    \caption{topo7}
    \label{subfig:topo7}
  \end{subfigure}%
  \begin{subfigure}[b]{0.5\textwidth}
    \centering
    \includegraphics[scale=0.7]{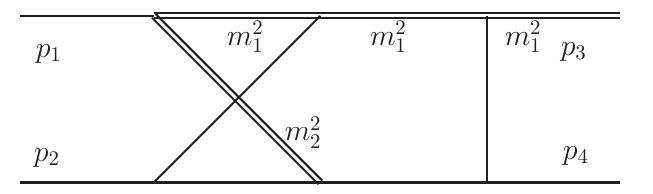}
    \caption{topo5}
    \label{subfig:topo5}
  \end{subfigure}%
  \\
  \begin{subfigure}[b]{0.5\textwidth}
    \centering
    \includegraphics[scale=0.7]{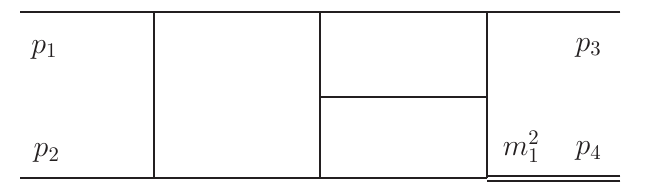}
    \caption{TennisCourt}
    \label{subfig:TennisCourt}
  \end{subfigure}%
  \begin{subfigure}[b]{0.5\textwidth}
    \centering
    \includegraphics[scale=0.7]{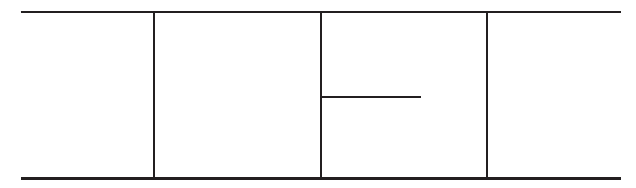}
    \caption{massless doublePentagon}
    \label{subfig:dp-0}
  \end{subfigure}%
  \\
  \begin{subfigure}[b]{0.5\textwidth}
    \centering
    \includegraphics[scale=0.7]{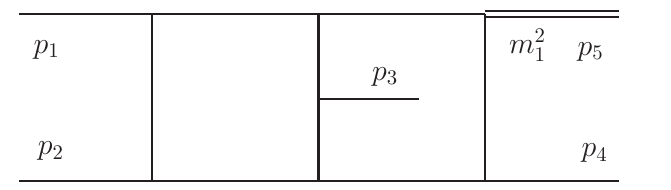}
    \caption{one mass doublePentagon}
    \label{subfig:dp-1}
  \end{subfigure}%
  \begin{subfigure}[b]{0.5\textwidth}
    \centering
    \includegraphics[scale=0.7]{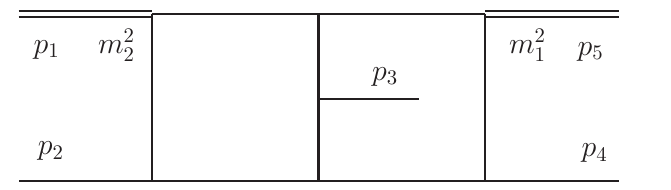}
    \caption{two mass doublePentagon}
    \label{subfig:dp-2}
  \end{subfigure}%
  \\
  \begin{subfigure}[b]{0.5\textwidth}
    \centering
    \includegraphics[scale=0.7]{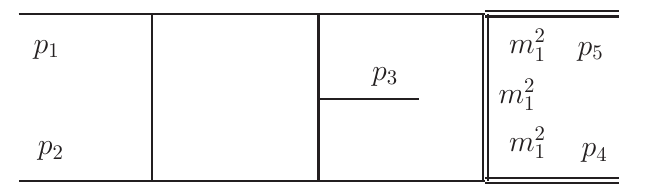}
    \caption{three equal mass doublePentagon}
    \label{subfig:dp-3}
  \end{subfigure}%
  \begin{subfigure}[b]{0.5\textwidth}
    \centering
    \includegraphics[scale=0.7]{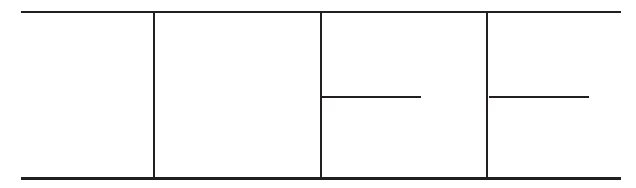}
    \caption{hex}
    \label{subfig:hex}
  \end{subfigure}%
  \caption{Illustrated are different families which are used to study the algorithms as implemented in \kirathr{}.
  Integral family \ref{subfig:topo7} and \ref{subfig:topo5}, are double-box topologies and are called \code{topo7} and \code{topo5}, respectively.
  Integral family \ref{subfig:TennisCourt} is a three-loop tennis-court topology, called \code{TennisCourt}.
  The integral family depicted in \ref{subfig:dp-0}-\ref{subfig:dp-3} belong to a double-pentagon topology and are called \code{doublePentagon}.
  Integral family \ref{subfig:hex} is a two loop pentagon-hexagon family with the name \code{hex}.
  The external momenta $p_i$, are explicitly depicted to avoid ambiguity.
  Variables $m_i$ are used to denote massive lines.}
  \label{fig:integrals}
\end{figure}

We provide multiple examples for two-loop and three-loop reductions in the \code{example} directory shipped with the source code, in which the option \code{truncate\_sp} as introduced in \cref{ssec:seeding} and the new equation selection discussed in \cref{ssec:selection} are demonstrated,
\begin{itemize}
  \item \code{double-box}: \code{topo7} and \code{topo5}, see \cref{subfig:topo7,subfig:topo5},
  \item \code{tennis-court}: \code{TennisCourt}, see \cref{subfig:TennisCourt},
  \item \code{double-pentagon}: \code{doublePentagon}, \code{doublePentagon-1-mass}, \code{doublePentagon-2-mass}, \code{doublePentagon-3-equal-mass}, see \cref{subfig:dp-0,subfig:dp-1,subfig:dp-2,subfig:dp-3},
  \item \code{pentagon-hexagon}: \code{hex}, see \cref{subfig:hex}.
\end{itemize}

To identify individual reduction problems we use the notation ${r_{\mathrm{max}}.s_{\mathrm{max}}.d_{\mathrm{max}}}$ in the following.
$r_{\mathrm{max}}$ stands for the maximum total number of positive indices, $s_{\mathrm{max}}$ for the maximum total number of negative indices, and $d_{\mathrm{max}}$ for the maximum total number of dots, cf.\ \cref{eq:rsd}.
All reductions are performed for all integrals from top-level sectors only.
We compare the new version \kirathr{} against the previous version \kiratwothr{}.
In all reductions \kira{}'s default integral ordering $1$ and the default permutation option $1$ are used, cf.\ \cref{ssec:order}.
All benchmarks are performed on a machine with two AMD EPYC 7702 64-core processors with 1\,TiB of main memory.

\subsection{Double-box topologies}

While \code{topo7} and \code{topo5} in \cref{subfig:topo7,subfig:topo5} are relatively simple two-loop topologies, they illustrate \kirathr{}'s efficiency for planar and non-planar diagrams with varying number of scales.
\code{topo7} has only one massive internal line and 31 master integrals, while \code{topo5} has multiple scales, 76 master integrals, and is non-planar.
\Cref{tab:double-box} shows that for \code{topo7} \kirathr{} generates 27 times less equations, finds 13 times less independent equations, and selects 4 times less equations.
\begin{table}[htb]
  \caption{
    The double-box integral families \code{topo7} and \code{topo5} are studied.
    They reduce to 31 and 76 master integrals, respectively.
  }
  \label{tab:double-box}
  \setlength{\tabcolsep}{5pt}
  \begin{center}
    \begin{tabular}{llrr@{\hspace{0.5em}}rr} 
      \toprule
      \multirow{5}{*}{\rotatebox{90}{problem}}
      &r.s.d& \multicolumn{2}{c}{7.5.0 (\code{topo7})} & \multicolumn{2}{c}{7.5.0 (\code{topo5})} \\[0.3em]
      \cmidrule(lr){3-4} \cmidrule(lr){5-6}
     &version& \multicolumn{1}{c}{\kiratwothr{}} & \multicolumn{1}{c}{\kirathr{}}& \multicolumn{1}{c}{\kiratwothr{}} & \multicolumn{1}{c}{\kirathr{}}  \\[0.3em]
     \cmidrule(lr){3-3} \cmidrule(lr){4-4} \cmidrule(lr){5-5} \cmidrule(lr){6-6}
      &\texttt{truncate\_sp}& - & l=3 & - & l=2 \\[0.3em]
      \midrule
      \multirow{5}{*}{\rotatebox{90}{generation}}
      &\#\,gen.& 615\,943 & 23\,030    & 774\,165    & 50\,804 \\
      &\#\,indep.& 216\,179 & 16\,109    & 325\,702    & 35\,443 \\
      &\#\,sel.& 38\,576 & 9\,473     & 126\,285    & 29\,740 \\
      &RAM\,[GiB]& 0.87   & 0.07   & 1.52 & 0.16  \\
      &$T_{\mathrm{gen}}$\,[min:s] & 0:23.3   & 0:02.2    & 1:15.5     & 0:04.7 \\
      \midrule
      \multirow{1}{*}{\rotatebox{90}{FF}} 
      & $T_{\pyred{}}$\,[s] & 0.16 & 0.04 & 1.4    & 0.36 \\
      \bottomrule
    \end{tabular}
  \end{center}
\end{table}
%
The ratio of independent equations to generated equations improves from 35\,\% to 70\,\% and the ratio of selected equations to generated equations from 6.2\,\% to 41\,\%, indicating a significantly improved strategy to only generate relevant equations.
This speeds up the generation step by a factor 11 and improves the memory footprint by a factor 12.
The solution time with \kira{}'s internal solver \pyred{} for a phase space point over a finite field improves by a factor of 4.0.

Similarly, 15 times less equations are generated for \code{topo5} of which now 59\,\% are selected instead of 16\,\% in \kiratwothr{}.
The generation is now 16 times faster, requires 9.5 less memory, while the solution with \pyred{} becomes 3.9 times faster.

\begin{figure}[ht]
  \centering
  \includegraphics[scale = 1]{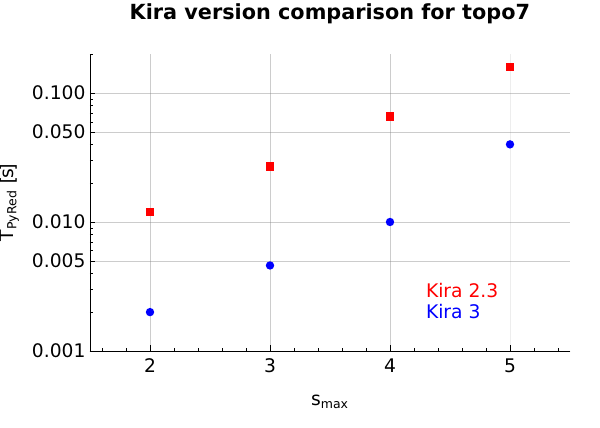}
  \caption{Scaling of the solution time $T_{\pyred{}}$ for the planar double-box \code{topo7} with \kirathr{} and \kiratwothr{}.}
  \label{fig:topo7-plot}
\end{figure}

In \cref{fig:topo7-plot} we compare the solution time over a finite field with \pyred{} for increasing values of $s_{\mathrm{max}}$ for \code{topo7}.
Compared to \kiratwothr{}, \kirathr{} can solve the system with $s_{\mathrm{max}} + 2$ in the same time, highlighting the improvements in the new version.
Overall, the improvements for these trivial examples are certainly welcome, but only the more complicated examples in the following subsections show the true power of the new version.



\subsection{Tennis-court topology}

\begin{figure}[ht]
  \centering
  \includegraphics[scale = 1]{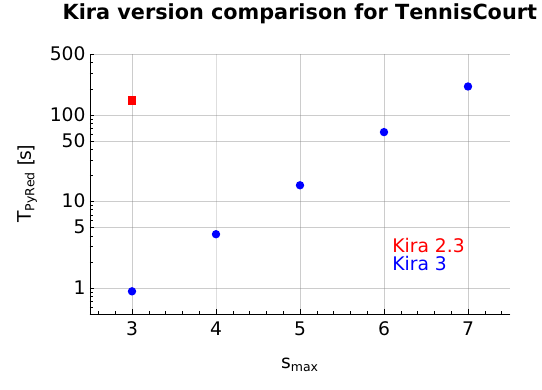}
  \caption{Scaling of the solution time $T_{\pyred{}}$ for the \code{TennisCourt} topology with \kiratwothr{} and \kirathr{}.}
  \label{fig:tennis-court-plot}
\end{figure}

As the first non-trivial example we study the three-loop \code{TennisCourt} topology depicted in \cref{subfig:TennisCourt}.
In \cref{tab:tennis-court} we show the reductions for \kiratwothr{} and \kirathr{} at the same complexity with $r_{\mathrm{max}} = 10$, $s_{\mathrm{max}} = 3$, and $d_{\mathrm{max}} = 0$.
\begin{table}[ht]
  \caption{
    The reductions are performed with different integral complexities for the three-loop integral family \code{TennisCourt}.
    The number of master integrals is 166.
  }
  \label{tab:tennis-court}
  \setlength{\tabcolsep}{5pt}
  \centering
  \begin{tabular}{llrr@{\hspace{1em}}r@{\hspace{1em}}r@{\hspace{1em}}r@{\hspace{1em}}r}
    \toprule
    \multirow{5}{*}{\rotatebox{90}{problem}}
    &r.s.d& \multicolumn{2}{c}{10.3.0} & \multicolumn{1}{c}{10.4.0} & \multicolumn{1}{c}{10.5.0} & \multicolumn{1}{c}{10.6.0} & \multicolumn{1}{c}{10.7.0} \\[0.3em]
    \cmidrule(lr){3-4} \cmidrule(lr){5-5} \cmidrule(lr){6-6} \cmidrule(lr){7-7} \cmidrule(lr){8-8}
    &version& \multicolumn{1}{c}{\kiratwothr{}} & \multicolumn{5}{c}{\kirathr{}} \\[0.3em]
    \cmidrule(lr){3-3} \cmidrule(lr){4-8}
    &\texttt{truncate\_sp}& - & l=7 & l=6 & l=5 & l=4 & l=3 \\[0.3em]
    \midrule
    \multirow{5}{*}{\rotatebox{90}{generation}} %
    &\#\,gen.& 8\,272\,762 & 387\,011 & 865\,442 & 2\,984\,878 & 9\,734\,618 & 27\,999\,684 \\
    &\#\,indep.& 4\,890\,067 & 282\,686 & 603\,856 & 1\,827\,965 & 5\,253\,869  & 13\,555\,275 \\
    &\#\,sel.& 2\,768\,144 & 115\,021 & 333\,055 & 1\,034\,357 & 2\,900\,250  & 7\,259\,008  \\
    &RAM\,[GiB]& 35.44 & 0.56 & 1.62 & 5.05 & 15.61 & 46.18 \\
    &$T_{\mathrm{gen}}$\,[min:s] & 41:36.3 & 0:28.7 & 1:09.4 & 3:23.0 &  11:43.8 & 46:30.2 \\
    \midrule
    \multirow{1}{*}{\rotatebox{90}{FF}}
    &$T_{\pyred{}}$\,[s] & 146     &  0.91    & 4.16 & 15.3 & 63  & 212 \\
    \bottomrule
  \end{tabular}
\end{table}
%
\kirathr{} generates about 21 times less equations and selects a system with about 24 times less equations.
This means that in this case the ratio of selected equations to generated equations drops slightly from 33\,\% to 30\,\%, staying on a acceptable level.
Furthermore, it requires about 63 times less main memory to set up the system and is about 87 times faster in doing so.
Finally, we see a performance improvement of a factor 160 in the reduction time on a finite field point.
To give glimpse what is now possible, we also show more complicated reductions by increasing $s_{\mathrm{max}}$ up to seven for \kirathr{}.
The tremendous effect is visualized in \cref{fig:tennis-court-plot} where a reduction of integrals with seven numerators with \kirathr{} is as efficient as a reduction of integrals with three numerators with \kiratwothr{}.
%

\subsection{Double-pentagon and pentagon-hexagon topologies}
\label{ssec:double-pentagon}

In \cref{tab:double-pentagon} we study the integral family \code{doublePentagon} depicted in \cref{subfig:dp-0}.
\begin{table}[ht]
  \caption{
    Comparison of \kiratwothr{} and \kirathr{} for the integral family \code{doublePentagon}.
    Both reduce to 108 master integrals.
  }
  \label{tab:double-pentagon}
  \centering
  \setlength{\tabcolsep}{5pt}
  \begin{tabular}{llrr@{\hspace{0.5em}}r@{\hspace{0.5em}}r@{\hspace{0.5em}}r}
    \toprule
    \multirow{2}{*}{\rotatebox{90}{problem}}
    &r.s.d& \multicolumn{2}{c}{8.5.0}  & \multicolumn{1}{c}{9.5.1}  & \multicolumn{1}{c}{13.5.5} \\[0.3em]
    \cmidrule(lr){3-4} \cmidrule(lr){5-5} \cmidrule(lr){6-6}
    &version& \multicolumn{1}{c}{\kiratwothr{}} & \multicolumn{3}{c}{\kirathr{}} \\[0.3em]
    \cmidrule(lr){3-3} \cmidrule(lr){4-6}
    &\texttt{truncate\_sp}& - & l=4 & l=4 & l=4 \\[0.3em]
    \midrule
    \multirow{5}{*}{\rotatebox{90}{generation}}
    &\# gen.& 16\,872\,564  & 73\,902   & 485\,560    & 29\,942\,304 \\
    &\# indep.& 4\,842\,650   & 53\,648   & 234\,344    & 7\,517\,845  \\
    &\# sel.& 1\,157\,381    & 41\,998   & 147\,756    & 2\,895\,663  \\
    &RAM\,[GiB]& 25.04  & 0.34 & 1.08 & 35.17 \\
    &$T_{\mathrm{gen}}$\,[min:s] & 33:46.0    & 0:08.8   & 0:45.8     & 44:46.6     \\
    \midrule
    \multirow{1}{*}{\rotatebox{90}{FF}}
    & $T_{\pyred{}}$\,[s] & 33.0 & 1.3  & 4.0 & 122.0 \\
    \bottomrule
  \end{tabular}
\end{table}
%
We find a significant improvement of the new version \kirathr{} compared to the previous version \kiratwothr{} in all metrics:
\kirathr{} generates 228 times less equations and selects 28 times less equations.
The ratio of independent equations to generated equations improves from 29\,\% to 73\,\% and the ratio of selected equations to generated equations from 6.9\,\% to 57\,\%, again highlighting that many irrelevant equations are no longer generated.
This translates to a reduction in main memory requirements by a factor of 74 and improves the runtime by a factor of 107.
Finally, the solution time for a phase space point over a finite field improves by a factor of 25.
In addition we demonstrate that \kirathr{} is also efficient when we include dots in addition to numerators.
Even the extreme case of five dots in the top-level sector is only of slightly higher complexity than the reduction without dots with \kiratwothr{}.

Due to their importance for phenomenology, two-loop $2 \to 3$ topologies have been studied extensively over the last decade and many ideas for efficient reductions have been developed.
We also tried the \code{doublePentagon} topology with two of those dedicated and more involved strategies: the block triangular form~\cite{Liu:2018dmc,Guan:2019bcx} as implemented in \code{Blade}~\cite{Guan:2024byi} and syzygy equations~\cite{Gluza:2010ws,Schabinger:2011dz,Ita:2015tya,Chen:2015lyz,Larsen:2015ped,Bohm:2017qme,Bohm:2018bdy,Bendle:2019csk,vonManteuffel:2020vjv} as implemented in \neatibp{}~\cite{Wu:2023upw,Wu:2025aeg}.
For the latter, we also tried the spanning cuts (spc.) strategy recently implemented in its latest version \code{1.1}.
For the block triangular system we take the system provided in \citere{Guan:2019bcx} 
and improved all of the coefficients with Horner's method.
Since we cannot compare the methods themselves properly with one single example, we did not necessarily tune the other methods to achieve their best performance.
Nonetheless, the corresponding measurements in \cref{tab:kira-blade-neatibp} allow us to make some interesting preliminary observations.
%
%
%
\begin{table}[ht]
  \centering
  \caption{
    Different strategies applied to the \code{doublePentagon} topology shown in \cref{subfig:dp-0} at the complexity 8.5.0.
    The numbers for syzygies with spanning cuts are summed over all cuts.
  }
  \label{tab:kira-blade-neatibp}
  \begin{tabular}{c|cccc}
    \hline
    & \kirathr{} & syzygies & syzygies (spc.)  & block triangular \\
    \hline
    $T_{\mathrm{gen}}$ & 8.8\,s & $\sim$\,hours & $\sim$\,hours & $\sim$\,hours \\
    \#\,equations & 41\,998 & 26\,106 & 23\,834 & 3\,497 \\
    \#\,terms & 734\,833 & 1\,498\,728 & 544\,355 & 388\,973 \\
    disk\,[MiB] & 3.7 & 46 & 8.6 & 23 \\
    $T_{\pyred}$\,[s] & 1.3 & 3.3 & 0.42 & 0.63 \\
    $T_{\ratracer{}}$\,[s] & 0.25 & 1.31 & 0.068 & 0.5 \\
    \hline
  \end{tabular}
\end{table}
%
While in this example the system generated by \kirathr{} contains more equations than the other systems, its equations consist of less terms on average and its size on disk in \kira{} readable format is actually smaller.
This can be explained by the fact that \kira{} produces standard \ac{ibp} equations with few terms and small algebraic coefficients, whereas the other strategies attempt to construct target oriented equations for the price of potentially longer equations and more complicated coefficients.

Despite these different characteristics, solving the systems with \pyred{} over finite fields results in comparable runtimes.
Whereas solving the \kira{} system is completely dominated by the numerical Gaussian elimination, substituting numbers into the coefficients is non-negligible for the other systems.

In addition to \kira{}'s internal solver \pyred{} we also solved the systems with the external solver \ratracer{}~\cite{Magerya:2022hvj}, which records and plays the trace of operations and performs further optimizations.
Again, we measure the time to solve the system for one phase space point over a finite field.
Compared to \pyred{}, all systems benefit, but to a different degree.
It is expected that different systems benefit from different solution strategies and solvers.
However, we did not try any other solver for this manuscript.
For example, \code{Blade} provides a solver specifically designed for the block triangular system~\cite{Guan:2024byi}.


As we have seen before, \kirathr{} is able to generate the system in a few seconds, whereas the other methods take a few hours.
In view of the many million, or even tens of million, sample points required for realistic applications, all of those generation times are typically negligible in the overall CPU budget.
However, it could become relevant if the reduction coefficients are constrained with more sophisticated algorithms, see e.g.\ \citeres{Abreu:2018zmy,Badger:2021imn,DeLaurentis:2022otd,Chawdhry:2023yyx,DeLaurentis:2025dxw}.

With the above, for this two-loop $2 \to 3$ topology, \kirathr{} now seems able to generate systems at a level of efficiency that was only achievable with dedicated strategies before.
It will be interesting to study whether this observation holds for other topologies.
For example, the three-loop $2 \to 3$ topology studied in~\citere{Liu:2024ont} was efficiently addressed by \neatibp{}. Testing whether \kirathr{} can reach the same level of efficiency is left for future explorations.

As a first glimpse in this direction, we study how \kirathr{} scales with more complicated topologies for state-of-the-art calculations.
In \cref{tab:double-pentagon one-mass} we first study the integral family \code{doublePentagon} when adding external and internal masses, see \cref{subfig:dp-0,subfig:dp-1,subfig:dp-2,subfig:dp-3}.
\begin{table}[ht]
  \caption{
    \kirathr{}'s performance for the more complicated double-pentagon topologies with external and internal masses from \cref{subfig:dp-0,subfig:dp-1,subfig:dp-2,subfig:dp-3} as well as the massless pentagon-hexagon topology shown in \cref{subfig:hex}.
  }
  \label{tab:double-pentagon one-mass}
  \centering
  \setlength{\tabcolsep}{5pt}
  \begin{tabular}{llrrrr@{\hspace{0.5em}}r}
    \toprule
    \multirow{2}{*}{\rotatebox{90}{problem}}
    &r.s.d& \multicolumn{4}{c}{8.5.0 (\code{doublePentagon}, l=4)} & \multicolumn{1}{c}{9.5.0 (\code{hex}, l=5)} \\[0.3em]\cmidrule(lr){3-6} \cmidrule(lr){7-7}
    &topology& massless & one-mass & two-mass & internal mass & massless \\[0.3em]
    & \#\,masters & 108 & 142 & 185 & 172 & 313 \\
    \midrule
    \multirow{5}{*}{\rotatebox{90}{generation}}
    & \#\,gen. & 73\,902 &  76\,045 & 77\,286 & 78\,480 & 339\,215 \\
    & \#\,indep. & 53\,648 & 55\,596 & 57\,034 & 57\,918 & 218\,502 \\
    & \#\,sel. & 41\,998 & 43\,827 & 45\,359 & 46\,231 & 156\,853 \\
    & RAM\,[GiB] & 0.34 & 0.37 & 0.46 & 0.43 & 1.88 \\
    & $T_{\mathrm{gen}}$\,[min:s] & 0:08.8 & 0:09.9 & 0:12.8 & 0:11.6 & 1:10.0 \\
    \midrule
    \multirow{2}{*}{\rotatebox{90}{FF}}
    & $T_{\pyred{}}$\,[s] & 1.35 & 1.6 & 2.5 & 2.1 & 16 \\
    & $T_{\ratracer{}}$\,[s] & 0.25 & 0.38 & 0.45 & 0.43 & 3.5 \\
    \bottomrule
  \end{tabular}
\end{table}
Even though these problems become significantly more complicated, be it in terms of the number of required sample points to obtain analytic reduction coefficients or the algebraic structure of the master integrals, there is only a mild increase in terms of main memory requirements and runtime when generating the system with \kirathr{} and sampling it for individual phase space points.
Only when adding an additional leg as in the topology \code{hex} depicted in \cref{subfig:hex}, there is a clear jump in complexity visible with \kira{}.

\section{Conclusions}
\label{sect:conclusions}

We presented the new version \code{3} of the Feynman integral reduction program \kira{} which brings enormous performance improvements and reductions of main memory requirements.
This is mainly driven by a refined seeding strategy, generating significantly less equations while still being able to fully reduce the target integrals, and an improved equation selection algorithm.
We quantified the performance improvements compared to \kiratwothr{} at the hand of several different topologies at two- and three-loop order and observed for a single two-loop $2 \to 3$ topology that the new version produces a system of equations on par with those generated by dedicated strategies.

Furthermore, we presented other new features such as the ability to add additional relations, sample the system numerically on chosen phase space points over finite fields, perform symbolic \ac{ibp} reductions, and check the master integral basis.

We already applied this new version to a wide range of different projects, see \citeres{Fael:2023tcv,Driesse:2024xad,Fael:2024vko,Driesse:2024feo,Abreu:2024fei}, and believe that many more ambitious projects in both high-energy and gravitational-wave physics become now accessible with the public release of \kirathr{}.
The source code is available on
\begin{center}
  \url{https://gitlab.com/kira-pyred/kira}
\end{center}
together with a more detailed changelog, installation instructions, and a wiki containing some best-practice guidelines.
Bug reports and feature requests can be submitted via the issue system.
In addition, a statically compiled version can be downloaded from
\begin{center}
  \url{https://kira.hepforge.org/}
\end{center}

While we have shown that the new seeding and selection strategies are vastly superior, further improvements seem realistic.
On the one hand, we discussed in \cref{ssec:selection} that the generated systems still contain irrelevant information which potentially could be filtered out with an improved selection algorithm, see e.g.\ \citeres{Bohm:2018bdy,Wu:2023upw}.
On the other hand, \citeres{vonHippel:2025okr,Song:2025pwy,Zeng:2025xbh} recently introduced the concept of \emph{priority functions} for seeding and applied machine learning techniques to find optimal distributions.
Such an approach could address both seeding and selection and it will be interesting to investigate this in more detail in the future.
In addition, new symmetry detection algorithms such as the one presented in \citere{Wu:2024paw} will allow us to uncover new relations between master integrals and could potentially lead to improved systems.

In this version we largely neglected the interpolation and reconstruction steps necessary to obtain analytic reduction coefficients.
While for small rational functions with only a few variables those are typically limited by the sampling, for complicated rational functions the inherent costs of those algorithms become a bottleneck and the achieved performance improvements diminish.
Thus, both technical and algorithmic improvements in \firefly{} might be necessary, e.g.\ following the ideas of \citere{Belitsky:2023qho,Chawdhry:2023yyx,Maier:2024djk,Smirnov:2024onl}.
On the other hand it might be worthwhile to sample the phase space for $2 \to 3$ and $2 \to 4$ processes instead of interpolating the reduction tables, as for example pursued in \citere{Agarwal:2024jyq}.
As presented in \cref{tab:double-pentagon one-mass}, the performance of the new version of \texttt{Kira} looks promising for this strategy.
At the hand of \ratracer{}~\cite{Magerya:2022hvj} we have seen that further speed ups are possible with technical improvements and better solution algorithms for linear systems, see also \citeres{Liu:2023cgs,Mangan:2023eeb}.
It might also be worthwhile to study GPU acceleration.

For some applications a traditional algebraic solving strategy can still be advantageous.
While \kira{}'s solver based on the computer algebra system \fermat{} already benefits from the improvements in this paper, further gains with other computer algebra systems seem possible~\cite{Mokrov:2023vva,Smirnov:2023yhb}.

\section*{Acknowledgments}

We thank Philipp Maierh\"ofer for his collaboration in early development stages and Samuel Abreu, Mathias Driesse, Pier Monni, Ben Page, Benjamin Sauer, and Yelyzaveta Yedelkina for testing the new version.
Furthermore, we thank Takahiro Ueda for providing a fix for the interface to \fermat{} and Yannick Klein and Vitaly Magerya for their contributions to \firefly{}.
We thank Xin Guan, Xiao Liu, Yan-Qing Ma, and Wen-Hao Wu for communication on \code{Blade}, Janko B\"ohm, Rourou Ma, Yingxuan Xu, and Yang Zhang for communication on \code{NeatIBP}, and all of them for comments on \cref{ssec:double-pentagon}.

The work of F.L.~was supported by the Swiss National Science Foundation (SNSF) under contract \href{https://data.snf.ch/grants/grant/211209}{TMSGI2\_211209}.
J.U.~is funded by the Deutsche Forschungsgemeinschaft (DFG, German Research Foundation) Projektnummer 417533893/GRK2575 “Rethinking Quantum Field Theory” and by the European Union through the European Research Council under grant ERC Advanced Grant 101097219 (GraWFTy). Views and opinions expressed are however those of the authors only and do not necessarily reflect those of the European Union or the European Research Council Executive Agency. Neither the European Union nor the granting authority can be held responsible for them.
Z.W.~is supported by The Hangzhou Human Resources and Social Security Bureau through The First Batch of Hangzhou Postdoctoral Research Funding in 2024. The authors acknowledge the computational resources provided by CERN.

Some of the Feynman diagrams were drawn with the help of Axodraw~\cite{Vermaseren:1994je} and JaxoDraw~\cite{Binosi:2003yf} and some with the help of \texttt{FeynGame}~\cite{Harlander:2020cyh,Harlander:2024qbn}.

\appendix


\bibliographystyle{JHEP}
\bibliography{bib}

\end{document}